\documentclass[prx,twocolumn,aps,epsf,showpacs,superscriptaddress,longbibliography,footinbib]{revtex4-1}
\usepackage[pdftex]{graphicx}

\usepackage{dcolumn}
\usepackage{bm}
\usepackage{epsfig}
\usepackage{latexsym} 
\usepackage{amsmath}
\usepackage{amssymb}
\usepackage{color}
\usepackage{array}
\usepackage{bbm}
\usepackage{hyperref}
\usepackage{color}
\usepackage{array}
\usepackage{cancel}

\usepackage[normalem]{ulem}
\usepackage{MnSymbol}

\usepackage{titlesec}
\titleformat{\paragraph}[runin]% runin puts it in the same paragraph
        {\bfseries}% formatting commands to apply to the whole heading
        {}% the label and number
        {0.0em}% space between label/number and subsection title
        {}% formatting commands applied just to subsection title
        [ -- ~]% punctuation or other commands following subsection title
\titlespacing*{\paragraph}{0pt}{4pt}{0pt}

\usepackage{booktabs}
\newcolumntype{C}[1]{>{\centering\arraybackslash}m{#1}}
\AtBeginDocument{
\heavyrulewidth=.08em
\lightrulewidth=.05em
\cmidrulewidth=.03em
\belowrulesep=.65ex
\belowbottomsep=0pt
\aboverulesep=.4ex
\abovetopsep=0pt
\cmidrulesep=\doublerulesep
\cmidrulekern=.5em
\defaultaddspace=.5em
}
\usepackage{booktabs}
\newcolumntype{R}[1]{>{\raggedleft\arraybackslash}p{#1}}
\AtBeginDocument{
\heavyrulewidth=.08em
\lightrulewidth=.05em
\cmidrulewidth=.03em
\belowrulesep=.65ex
\belowbottomsep=0pt
\aboverulesep=.4ex
\abovetopsep=0pt
\cmidrulesep=\doublerulesep
\cmidrulekern=.5em
\defaultaddspace=.5em
}

\newcommand{\<}{\langle}
\newcommand{\e}{\varepsilon}
\newcommand{\up}{\uparrow}
\newcommand{\down}{\downarrow}
\renewcommand{\>}{\rangle}
\renewcommand{\(}{\left(}
\renewcommand{\)}{\right)}
\renewcommand{\[}{\left[}
\renewcommand{\]}{\right]}
\renewcommand{\d}{\partial}

\newcommand{\Z}{\mathbb{Z}}

\begin{document}
\title{A matrix product operator approach to non-equilibrium Floquet steady states}
%\title{Matrix-product-operator approach for steady states of Floquet-Dissipative many-body systems}

\author{Zihan Cheng}
\affiliation{Department of Physics, University of Texas at Austin, Austin, TX 78712, USA}

\author{Andrew C. Potter}
\affiliation{Department of Physics, University of Texas at Austin, Austin, TX 78712, USA}
\affiliation{Department of Physics and Astronomy, and Quantum Matter Institute, University of British Columbia, Vancouver, BC, Canada V6T 1Z1}

\begin{abstract}

We present a numerical method to simulate non-equilibrium Floquet steady states of one-dimensional periodically-driven (Floquet) many-body systems coupled to a dissipative bath, called open-system Floquet DMRG (OFDMRG). This method is based on a matrix product operator ansatz for the Floquet density matrix in frequency-space, and enables access to large systems beyond the reach of exact master-equation or quantum trajectory simulations, while retaining information about the periodic micro-motion in Floquet steady states.
An excited-state extension of this technique also allows computation of the dynamical approach to the steady state on asymptotically long timescales. We benchmark the OFDMRG approach with a driven-dissipative Ising model, and apply it to study the possibility of dissipatively stabilizing pre-thermal discrete time-crystalline order by coupling to a cold bath. 

%This method is a direct extension of the conventional density-matrix renormalization group and hence feature the ability to access large system size. 
%Also, the inclusion of the frequency-space allows the descripation of micromotion in the Floquet steady states. Our numerical implementation on both dissipative driving-Ising model and dissipative discrete time crystal model demonstrates the efficiency of the method for finding Floquet steady states. We also adapt excited state techniques to compute the dynamics
%We show that by adding proper penalizing term one can also compute
\end{abstract}
\maketitle

%% Introductory text %%
Controlling quantum systems with time-periodic (Floquet) external driving fields offers a powerful toolkit to engineer interactions, symmetry-breaking, and topology that are not present in the un-driven system~\cite{Oka2018}.  
Floquet driving can also produce \emph{intrinsically non-equilibrium} phenomena such as dynamical phases like time crystals and Floquet topological phases~\cite{Harper2019,Else2020}, with properties that would be impossible in static equilibrium. 
However, for isolated systems, persistent energy absorption from external ordinarly produces runaway heating to a featureless state\cite{Dalessio2014,Lazarides2014} that is locally indistinguishable from an infinite temperature ensemble.
Thus, to stabilize dynamical phases in closed Floquet systems, one usually considers systems with many-body localization~\cite{Lazarides2015} that fail to thermalize, or work in a pre-thermal regime~\cite{Abanin2015,Abanin2017a,Abanin2017b,Bukov2015,Else2017,Eckardt2015,Mizuta2019} where Floquet states can live up an exponentially-long timescale $\tau_\text{heat}\sim e^{\Omega/\Lambda}$ in the ratio of driving frequency $\Omega$, to the local bandwidth, $\Lambda$. 
Both of these approaches have substantial limitations. 
First, MBL requires synthesizing strong disorder, and is fundamentally incompatible with many interesting phenomena such as non-Abelian symmetries and anyons~\cite{potter2016symmetry}, Goldstone modes~\cite{banerjee2016variable}, long-range interactions and (at least as a matter of principle if not practice) in dimensions higher than one~\cite{DeRoeck2017}. 
Second, no experimental system is truly isolated from its environment, which restricts MBL-protected order to transient times. 
Realizing pre-thermal \emph{quantum} phases requires preparing a low-temperature state of the pre-thermal Hamiltonian which is typically hard to even calculate, let alone prepare its ground-state (e.g. adiabatic state-preparation generally fails in Floquet settings~\cite{Weinberg2017}).

%A natural solution, familiar from solid-state physics, would be to 
Experience from solid-state physics, it is natural to look to dissipation from a cold bath to cool a Floquet system close to its pre-thermal ground-state. 
For fast, weakly-heating drives, rigorous bounds on pre-thermalization~\cite{Abanin2015,Abanin2017a,Abanin2017b,Bukov2015,Else2017,Eckardt2015,Mizuta2019} establish a large separation of time scales between the drive period $\tau=2\pi/\Omega$, and the heating time $\tau_\text{heat}$.
This suggests an ample range of parameter space to couple the system to a bath weakly enough to avoid disrupting the interesting Floquet dynamics, while cooling towards the pre-thermal ground-state at a rate much higher than the drive-induced heating.
On the other hand, coupling a system to a bath can enhance drive-induced heating, by broadening spectral lines in the system to enable off-resonant drive-induced excitations that cause the system to heat~\cite{Rudner2020}.
To explore the balance between these competing processes and establish whether dissipation can stabilize dynamical orders in an appropriately designed range of drive, bath, and system-bath coupling parameters, it requires a controlled calculation method that can simultaneously treat strong driving, interactions, and open system dynamics.

However, solving the long-time non-equilibrium steady state (NESS) of a generic Floquet-Lindblad equation~\cite{Haddadfarshi2015,Dai2016,Ikeda2020,Ikeda2021} (FLE) is a challenging task, even for one-dimensional systems. 
Similar to solving Schrodinger equation, the cost of exact treatment grows exponentially with respect to the system size, but with a \emph{double}(!) exponent due to simulating density matrices rather than pure states. 
Quantum trajectory sampling methods~\cite{Molmer1993, Dalibard1992, Dum1992} reduce the memory cost, but may incur exponential-in-system-size sampling overheads.

In one dimension ($1d$), matrix product states (MPS) and operators (MPO) provide an effective way of representing systems with limited spatial entanglement -- a class that includes not only ground-states of gapped systems~\cite{Schollwock2005} but also thermal mixed-states~\cite{Wolf2008}.
One class of MPO approaches~\cite{Verstraete2004,Zwolak2004,Vidal2004,White2004}, in combination with time-evolving block decimation (TEBD) methods, allows studying the NESS via long-time dynamics. 
Such real-time approaches can suffer from the long relaxation time to the NESS, for example in the presence of long-time hydrodynamic tails, and weakly-dissipative systems may also feature a rapid growth of entanglement in the transient regime that cannot be captured by a low bond-dimension MPO~\cite{Cui2015,Mascarenhas2015,Weimer2021}, presenting a short-time barrier to accessing the NESS through time-evolution.

%\blue{In $1d$, matrix product states (MPS) and operators (MPO) provide an effective way of representing systems with limited spatial entanglement in $1d$ -- a class that includes not only ground-states of gapped systems~\cite{Schollwock2005} but also thermal mixed-states~~\cite{Wolf2008}.One class of MPO approaches~\cite{Verstraete2004,Zwolak2004,Vidal2004,White2004} involve performing real time-evolution using open-system time-evolving block decimation (TEBD) methods, can be effective for strong dissipation. However, for weakly dissipative systems the approach to the steady state can occur through a transient regime with rapid growth of entanglement that cannot be captured by a low bond-dimension MPO~\cite{Cui2015,Mascarenhas2015,Weimer2021}.}

To overcome these limitations, for time-independent systems, recent works~\cite{Cui2015, Mascarenhas2015} directly target an MPO representation of a NESS that is variationally optimized through density matrix renormalization group (DMRG) type methods~\cite{White1992}.
%The density matrix renormalization group (DMRG)~\cite{White1992} based on the matrix product state (MPS)~\cite{Verstraete2006,Hastings2007} ansatz has shown remarkable success in describing ground states of one-dimensional gapped local Hamiltonians.  It is natural to extend the MPS from pure to mixed states, where matrix product operators (MPOs) can be used as an ansatz for density matrices~\cite{Verstraete2004, Zwolak2004}, enabling progress in describing the dynamics and steady state properties of open quantum systems~\cite{Zwolak2004, Vidal2004, White2004, Mascarenhas2015, Cui2015}. Among them, TEBD-like algorithms~\cite{Zwolak2004, Vidal2004, White2004} have been implemented to evolve an initial state in the real time, which automatically fits for the Floquet system. However, this dynamical approach is not always efficient to targeting the steady states, since the low dissipative rate of the system could cause is long relaxation time. Also, the intermediate state in the evolution may not obey the area law, which can severely affect the convergence~\cite{Weimer2019}. To overcome these limitations, recent works directly target the steady states using variational searching or imaginary time evolution. 
 %
In this paper, we extend this technique to open Floquet systems. The central idea will be to reduce the time-dependent Floquet problem to an effective time-independent one in an extended (frequency) space. Frequency-space methods are widely used in various analytic and numerical approaches to Floquet problems~\cite{rudner2020floquet}. Here we adapt this representation in a form convenient for performing MPS calculations.
 Importantly, the method retains information not only about the NESS at stroboscopic times, but also the micro-motion within a period, which can be required to observe certain dynamical phases, such as Floquet topological insulators and symmetry-protected topological phases~\cite{Harper2019}. 
 We also benchmark this method with a driven-dissipative Ising model and use it to explore the dissipative stabilization of a discrete time-crystal (DTC) by coupling it to a cold bath.
% We benchmark this method using a simple periodically-driven and dissipative Ising model, and establish that it can produce high precision computation of the observables and micro-motion of the NESS of this model for small sizes that can be compared to exact calculations, but also can readily extend to much larger systems.
% Next, we apply this numerical technique to explore the dissipative stabilization of a pre-thermal discrete time-crystal (DTC) in a periodically kicked Ising chain coupled to an ohmic bath. This model is similar to one recently explored in trapped-ion experiments~\cite{}, but importantly, we consider geometrically local interactions which require preparing the system in its pre-thermal ground-state to observe long-ranged DTC order. We show that under appropriate conditions on the drive-frequency, system-bath coupling, and bath spectral characteristics, that one can achieve a Floquet Gibbs state in a rotating frame, with a temperature close to that of the bath, and stabilize long-lived and spatially extended DTC order via dissipation.
\begin{figure}[t]
	\includegraphics[width=0.8\columnwidth]{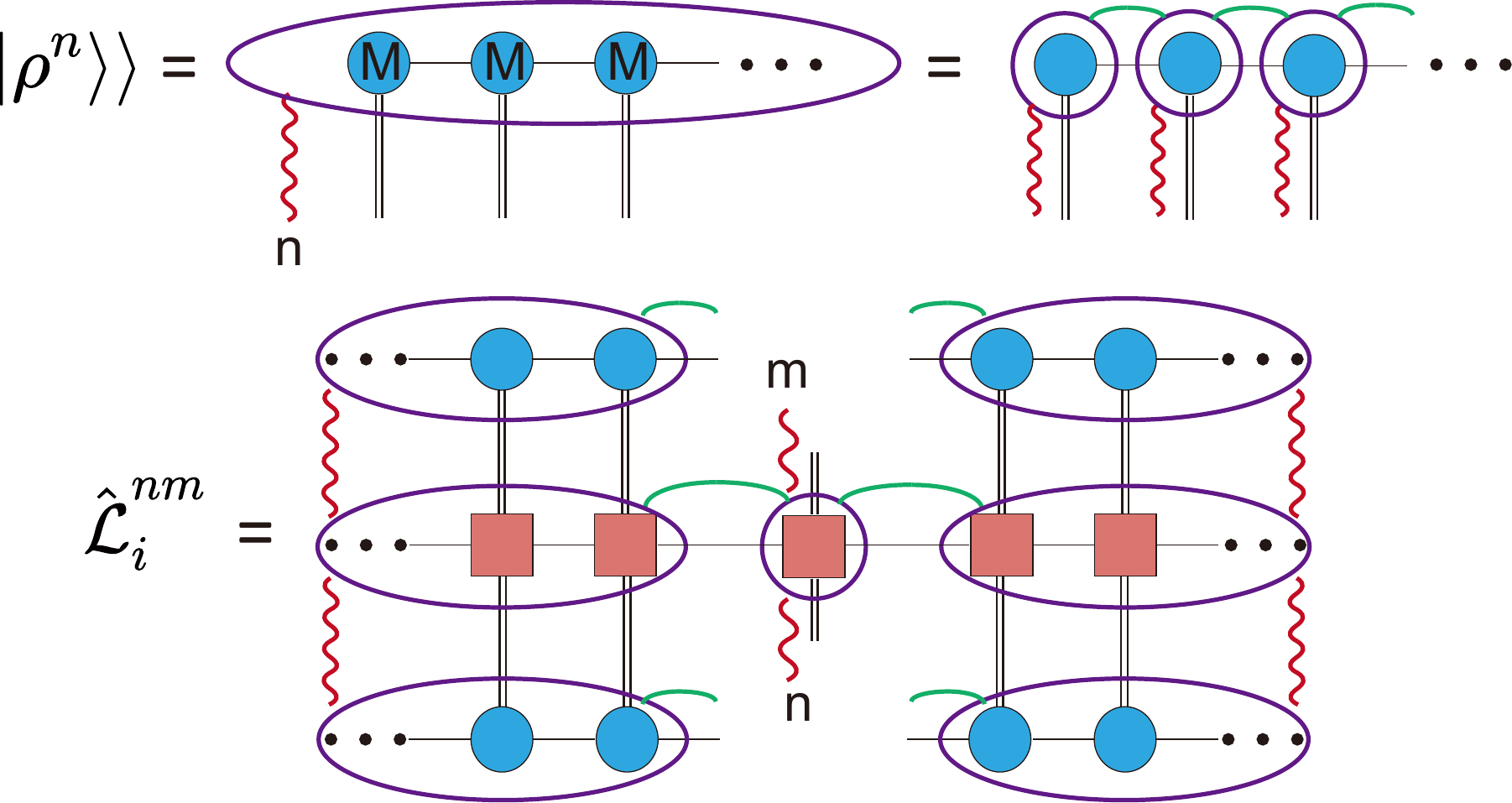}
	\caption{{\bf Graphical representation --} of vectorized MPOs $|\rho^n\rrangle$ (upper) and effective local Lindbladian $\hat{\mathcal{L}}^{nm}_i$ (lower) in frequency-space. Blue circles and black single/double lines respectively represent the tensor and bond/physical indices in the Hilbert space, while the purple circle and red wavy lines represent the tensor and Fourier indices in frequency-space. Green lines represent the virtual bond space where each tensor is block diagonal in $n$.
	%The second equal sign shows the frequency-space index can be equivalently distributed to each a local tensor $M^n$ with additional identical-$n$-component restriction, represented by green lines, which allows the local variable searching with respect to each $M^n$. (lower) Graphic representation of effective extended Lindbladian $\hat{\mathcal{L}}^{nm}_i$ with respect to site $i$.
	}
	\vspace{-0.2in}
	\label{fig:graphicrepresentation}
\end{figure}

%% Method %%
\paragraph{Frequency-space MPO representation}
Consider the evolution of the density matrix $\rho(t)$ of a periodically driven $1d$ quantum system coupled to a Markovian bath described by the Floquet-Lindblad equation (FLE): 
%$\d_t\rho = \mathcal{L}(t)\[\rho\]=-i\[H(t),\rho\]+\sum_\alpha\(L_\alpha(t)\rho L_\alpha^\dagger(t)-\frac{1}{2}\left\{L^\dagger_\alpha(t) L_\alpha(t), \rho\right\}\),$~\cite{}:
\begin{align}\label{eq:FLE}
\d_t\rho =& \mathcal{L}(t)\[\rho\]=-i\[H(t),\rho\] \nonumber\\&+\sum_\alpha\(L_\alpha(t)\rho L_\alpha^\dagger(t)-\frac{1}{2}\left\{L^\dagger_\alpha(t) L_\alpha(t), \rho\right\}\),
\end{align}
where $H(t+\tau)=H(t)$ and $L_\alpha(t+\tau)=L_\alpha(t)$ are respectively the periodic Hamiltonian and jump operators.

Floquet's theorem enables one to write solutions to the FLE in terms of quasi-eigenmodes of the Lindbladian $\mathcal{L}(t)$ as: $\rho(t) = \sum_n \rho^n e^{-\lambda t} e^{in\Omega t}$, where $\lambda$ is the (complex) quasi-eigenvalue and $\Omega = 2\pi/\tau$ is the driving frequency (See Appendix~\ref{appendix:floquettheorem} for details). 
Inserting this expression into Eq.~\ref{eq:FLE}, reduces the time-dependent FLE into an effectively time-independent equation: $\hat{\mathcal{L}}[\hat\rho]= -\lambda \hat\rho$ for extended $\hat\rho = \sum_n \rho^n\otimes |n\rrangle$ residing in an enlarged (frequency) space $\mathcal{H}^2\times \Z$ (intuitively, the extra $\Z$ factor keeps track of how many drive quanta the system has absorbed or released), where the extended Lindbladian is given by:
\begin{align}\label{eq:frequencyLindbladian}
	%-\lambda\rho^n= -in\Omega\rho^n-i\[H^{n-m}, \rho^m\] +\sum_\alpha D_\alpha^{nm}[\rho_m]
	\hat{\mathcal{L}}^{nm}[\rho^m] &= -in\Omega\rho^n\delta_{nm}-i\[H^{n-m}, \rho^m\] +\sum_\alpha D_\alpha^{nm}[\rho^m],
	\nonumber\\
	D^{nm}_\alpha[\rho^m] &= 
	L_\alpha^{n-k}\rho^mL_\alpha^{\dagger, k-m}-\frac{1}{2}\left\{L_\alpha^{\dagger,n-k}L_\alpha^{k-m}, \rho^m\right\},
\end{align}
where $H^n$, $L^n_\alpha$ are Fourier coefficients of $H$ and $L_\alpha$ with frequency $n\Omega$ respectively, and throughout this paper repeated Fourier indices are implicitly summed.
%In practice, we will truncate the infinite $\Z$ direction in frequency-space, which is valid when the linear ``potential" $-in\Omega$ leads to localization near $n=0$, characterized by rapid decay of $|\rho^n|/|\rho^0|$ with $n$(See Appendix~\ref{appendix:warmupconvergence}).

We are targeting models with high-frequency drives and weak-system bath couplings to model whether a system can be cooled close to a pre-thermal ground-state. Here, we expect $\rho^0$ to be approximately thermal, and hence exhibit an area-law operator entanglement~\cite{Wolf2008} permitting efficient representation as an MPO. 
We further assume that, at high frequencies, the linear potential $-in\Omega$ in frequency-space leads to localization near $n=0$ characterized by rapid decay of $|\rho^n|/|\rho^0|$ with $n$ (See Appendix~\ref{appendix:warmupconvergence} for convergence check), so that we can cut off the infinite frequency index beyond $|n|=N_c$, and that each $\rho^n$ has a low bond-dimension MPO representation  $\forall n$. 
The validity of assumptions can be checked \emph{a posteriori}. 
We note that the Fourier index $n$ can be regarded either as a global index, or distributed to each MPO tensor as an auxiliary label to the virtual bond space where each tensor is block diagonal in $n$ (see Fig.~\ref{fig:graphicrepresentation} for a graphical representation).
It is further convenient to vectorize the density matrices $\rho^n\rightarrow |\rho^n\rrangle$ using the Choi isomorphism $|\psi\>\<\phi|\rightarrow |\psi\otimes \phi\rrangle$, so that we regard the MPO as an MPS with squared physical dimension:
 %$|\rho\rrangle = \displaystyle\bigoplus_{|n|\leq N_c} |\rho^n\rrangle$ with:
\begin{align}
|\rho^n\rrangle = \sum_{\{\mu_i\}}M^n_{\mu_1}\dots M^n_{\mu_L} |\mu_1\dots \mu_L\rrangle,
\end{align}
where each $M^n$ is a $d^2\times \chi\times\chi$ tensor, $d$ is the onsite Hilbert space dimension, $\mu_i\in\{1\dots d^2\}$ labels a basis of physical states for the vectorized density matrix, $i=1\dots L$ label sites of the $1d$ chain, and $\chi$ is the bond dimension.

After the vectorization, $\hat{\mathcal{L}}^{nm}$ in Eq.(\ref{eq:frequencyLindbladian}) becomes a linear operator acting on $ \vert\rho^m\rrangle$, which can be similarly represented in an MPO form with two Fourier components $n,m$:
\begin{align}
	\hat{\mathcal{L}}^{nm}=\sum_{\{\mu_i,\nu_i\}}v^LW^{nm}_{\mu_1\nu_1}\cdots W^{nm}_{\mu_N\nu_N}v^R\vert \mu_1\cdots\mu_N\rangle\langle\nu_1\cdots\nu_N\vert,
\end{align}
where each $W^{nm}$ is a $d^2\times d^2\times \chi_O\times \chi_O$ tensor, $\chi_O$ are the operator bond dimension, and $v^{L,R}$ impose boundary conditions.

\paragraph{Open-system Floquet DMRG (OFDMRG)}%Variational preparation of the NESS}
In conventional MPS-DMRG for closed systems, one minimizes the variation energy $\<\psi|H|\psi\>$ for each local MPS tensor, which relies heavily on Hermiticity of $H$. 
A natural generalization~\cite{Cui2015} to open systems would be to minimize $\llangle \rho|\mathcal{L}^\dagger\mathcal{L}|\rho\rrangle$, however, the MPO for $\mathcal{L}^\dagger\mathcal{L}$ has square of the bond-dimension of that for $\mathcal{L}$, adding significant overhead~\cite{Mascarenhas2015}.
In an alternative approach~\cite{Mascarenhas2015}, instead of variationally searching for the local MPS, one can solve the zero eigenvector for the local effective Lindbladian ${\mathcal{L}}_i$ obtained by contracting all indices for $\llangle\rho\vert \mathcal{L}\vert\rho\rrangle$, except those for a single site $i$, so that sites $j\neq i$ form an environment for site $i$.

Here we adapt this approach directly to the frequency-space representation of $\rho$ and $\hat{\mathcal{L}}$, seeking to approximately prepare the NESS satisfying $\hat{\mathcal{L}}^{nm}_i|\rho^m\rrangle=0$ by sweeping through a sequence of local eigenvalue problems for $M^m_{\mu_i}$ (see Fig.~\ref{fig:graphicrepresentation}), using an implicitly restarted Arnoldi method based non-Hermitian eigensolver implemented in the ARPACK library~\cite{Arpack}.
Working in frequency-space requires imposing additional constraints on the solutions. 
Physical states satisfy ${\rm Tr}\rho(t)=\llangle \mathbb{I}\vert\rho(t)\rrangle=1\forall t$, which demands ${\rm Tr}\rho^n=\llangle \mathbb{I}\vert\rho^n\rrangle=\delta_{n0}$ where $|\mathbb{I}\rrangle$ is the maximally mixed state.
We enforce this condition by penalizing violations by modifying how the extended Lindbladian acts on vectors as $\hat{\mathcal{L}}\rightarrow \hat{\mathcal{L}}'$ with:
\begin{align}
	\hat{\mathcal{L}}'^{nm}|\rho^m\rrangle=&
	\(\hat{\mathcal{L}}^{nm}-
	P_0| \mathbb{I}\rrangle\llangle \mathbb{I}|(1-\delta_{n0})\delta_{nm}\)|\rho^m\rrangle\nonumber\\&-P_1\exp\(-|{\rm Tr}\rho^0|^2/\delta^2\)|\rho^{n}\rrangle,
\end{align}
where $P_0, P_1, \delta$ are penalty parameters. In practice, we start with several warm-up sweeps with proper penalty parameters ($P_0=P_1=1000$ and $\delta=0.01$ in our implementation) to avoid local minima violating the trace constraint, and then remove the penalty for further DMRG sweeping. (See Appendix~\ref{appendix:warmupconvergence} and~\ref{appendix:positivity} for discussion on convergence and positivity of density matrices)

\paragraph{Dynamical approach to the NESS}
The MPO-based method can be naturally extended to solve long-lived decaying modes of Floquet Lindbladian, with ${\rm Re}\lambda>0$, by a similar approach to excited state DMRG~\cite{Stoudenmire2012}. 
To explore this, we first review some basic properties of the (extended) Lindbladian: (i) the Lindbladian has a bi-orthornormal basis, where left and right eigenvectors are defined by  $\hat{\mathcal{L}}\vert\rho^R_\alpha\rrangle = \lambda_\alpha\vert\rho^R_\alpha\rrangle $ and $\hat{\mathcal{L}}^\dagger\vert\rho^L_\alpha\rrangle = \lambda^\ast_\alpha\vert\rho^L_\alpha\rrangle $ and satisfy the orthorgonal relations $\llangle \rho^L_\alpha\vert\rho^R_\beta\rrangle=\delta_{\alpha\beta}$; 
(ii)  The corresponding eigvenvalues $\{\lambda_{\alpha=0,1,\dots}\}$ can be sorted as $0 = \lambda_0>{\rm Re}\lambda_1\geqslant{\rm Re}\lambda_2\geqslant\cdots$ (we assume that the zero eigenvalue is not degenerate in the following discussion). 
In particular $\vert\rho^L_0\rrangle=\vert\mathbb{I}\rrangle$ due to the trace preservation of Lindblad operator; (iii) The complex eigenvalues must occur in a pair of complex conjugate since when $\rho$ is an eigenvector, $\rho^\dagger$ is also an eigenvector.

Based on properties of the Lindbladian and in analogy to the Hamiltonian case~\cite{Stoudenmire2012}, one can define $\hat{\mathcal{L}}_1=\hat{\mathcal{L}}-w\vert\mathbb{I}\rrangle\llangle\mathbb{I}\vert$ $(\hat{\mathcal{L}}^{\dagger}_1=\hat{\mathcal{L}}^\dagger-w\vert\rho_{ss}\rrangle\llangle\rho_{ss}\vert)$ where $w$ is the penalty energy for the vector not orthogonal to the zeroth left (right) eigenvector. For large enough $w$,  the solved eigenvalue with largest real part will give the first right (left) eigenvector $\vert\rho^R_1\rrangle$ ($\vert\rho^L_1\rrangle$). 
In principle, this procedure can be done recursively to the $n^{\rm th}$ eigenvector by adding $n$ projectors, however for the pair of eigenvectors whose eigenvalues are in complex conjugate pairs $\lambda=a\pm ib$, they cannot be distinguished by their real part.
Thus, we focus only on the first decaying mode by targeting the largest real part of eigenvalues, which dominates the approach to the steady state at long times.

\begin{figure}[t]	
	\includegraphics[width=1.05\columnwidth]{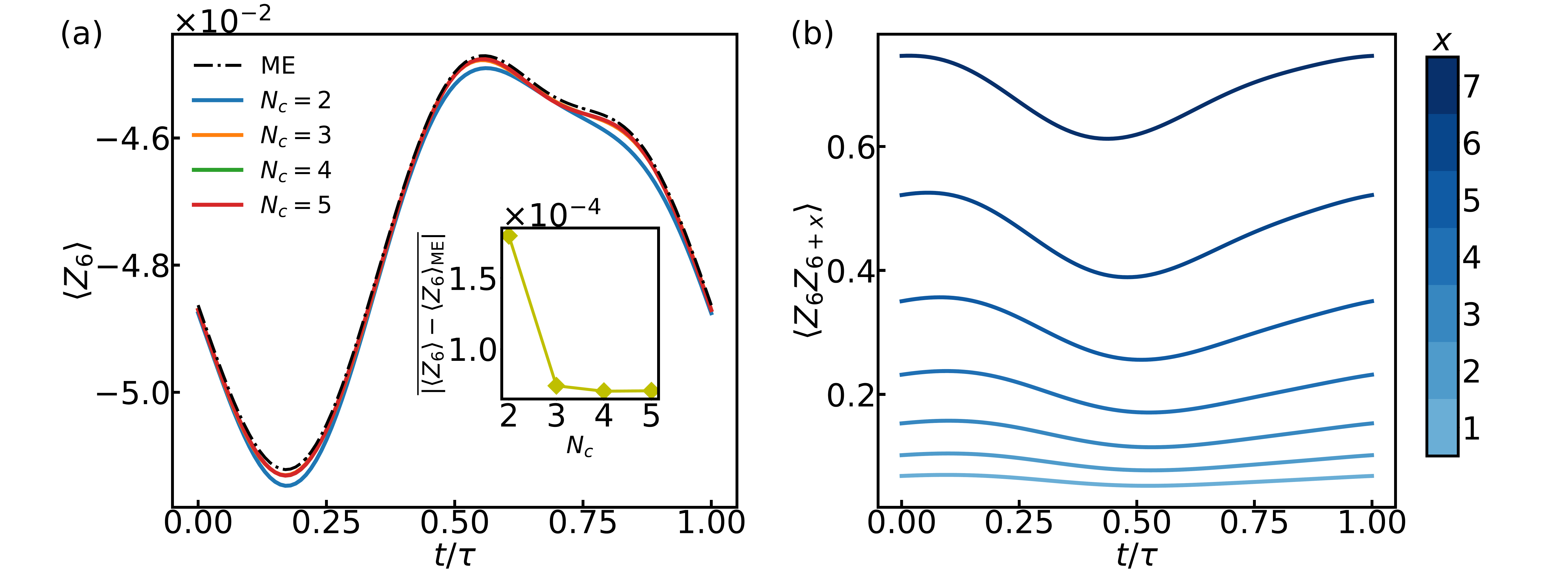}
	\caption{{\bf NESS of driven-dissipative Ising chain -- } with $\(J, h, g, \gamma, \omega\)=\(1.0, 0.5, 1.0, 1.0, 5.0\)$. (a) Time-dependent expectation values of magnetization $\langle Z_6\rangle$ for a system size $L=11$, with $\chi=36$,
	%, and $\(J, h, g, \gamma, \omega\)=\(1.0, 0.5, 1.0, 1.0, 5.0\)$ 
	compared with the master equation evolution result. 
	The period-averaged error (inset) 
	%in local observables, $\overline{\vert\langle Z_5\rangle-\langle Z_5\rangle_{\rm ME}\vert}$ averaged over one period, 
	decays rapidly with $N_c$ to the numerical accuracy of the eigensolver.
	%The inset of $a$ shows the error $\overline{\vert\langle Z_5\rangle-\langle Z_5\rangle_{\rm ME}\vert}$ averaged over one drive period versus $N_c$. 
	(b) Spatial correlations $\langle Z_6Z_{6+x}\rangle$ for a larger chain with $L=21$, using $(N_c, \chi)=(4, 20)$.
	%,  and $\(J, h, g, \gamma, \omega\)=\(1.0, 0.5, 1.0, 1.0, 5.0\)$.}
	}
	\label{fig:drivenising}
	\vspace{-0.2in}
\end{figure}

%\begin{figure}[t]
%	\includegraphics[width=0.9\columnwidth]{figcorrelationt.pdf}
%	\caption{Spatial correlations of the dissipative driving-Ising model for $N=21$, $d=20$, $N_c=4$, $J=-1.0$, $h=0.5$, $g=1.0$, $\omega=5$ and $\gamma=1.0$. }
%	\label{fig:correlationt}
%	\vspace{-0.2in}
%\end{figure}

\paragraph{Benchmark: driven-dissipative Ising chain}
We first benchmark our OFDMRG method in a driven-dissipative Ising model on a length $L$ spin-1/2 chain with Pauli operators $\{X_i,Y_i,Z_i\}$ for sites $i=1\dots L$ with Hamiltonian:
\begin{align}
	H(t) = \sum_i \[p(t)\(-JZ_iZ_{i+1}+hZ_i\)+q(t)gX_i\],
\end{align}
where $p(t)=\(1-\sin \omega t\)/2$, $q(t)=\(1+\sin \omega t\)/2$,
and time-independent majority-rule jump operators
\begin{align}
	L_i=%&\sqrt{\gamma}\(X_i+\frac{1}{2}iY_iZ_{i+1}+\frac{1}{2}iZ_{i-1}Y_i\)\nonumber\\
	%=&
	\sqrt{\gamma}(
	2|\up\up\up\>\<\up\down\up|%+2|\down\down\down\>\<\down\up\down|
	+|\up\up\down\>\<\up\down\down|
	+  |\up\down\down\>\<\up\up\down| +(\up\leftrightarrow\down)
	).
%	\nonumber\\&
%	+ |\up\down\down\>\<\up\up\down|+|\down\up\up\>\<\down\down\up|+|\down\down\up\>\<\down\up\up|)
\end{align}
To compare our method with the exact evolution of Lindblad master equation implemented in QUTIP~\cite{Johansson2013}, we simulate a chain with array length $L = 11$. 
We find excellent convergence in the central magnetization $\langle Z_6\rangle$ to the exact solution with increasing frequency-space cutoff $N_c$, achieving residual error $\sim 10^{-4}$ for $N_c\sim5$ that is consistent with residual error in the zero-eigenvalue solver of OFDMRG and the order of magnitude of Schmidt components at the bond-dimension cutoff (See Appendix~\ref{appendix:warmupconvergence}).  
%Fig.\ref{fig:drivenising}(a) and Fig.\ref{fig:drivenising}(b) show the steady-state evolution of magnetization $\langle Z_5\rangle$ and nearest-neighboring correlation  $\langle Z_5Z_6\rangle$ obtained by solving the Lindblad master equation and our OFDMRG method. As increasing dimension cutoff $N_c$ of the frequency-space from $2$ to $5$, OFDMRG results show a convergence  to the exact result of master-equation solution. As shown in Fig.\ref{fig:drivenising}(c) and Fig.\ref{fig:drivenising}(d), the time-dependent difference between OFDMRG results and exact evolution is at the order of $10^{-4}$, which is consistent with residual error in the zero-eigenvalue solver of OFDMRG and the order of magnitude of Schmidt component at the bond dimension cutoff (See Appendix B).  
The OFDMRG method also extends straightforwardly to larger systems with polynomial-in-$L$ scaling. 
For example, in Fig.~\ref{fig:drivenising} we show spatial correlations for a size $L=21$, which would require enormous computational resources to compute exactly.

\begin{figure}[t]
	\includegraphics[width=\columnwidth]{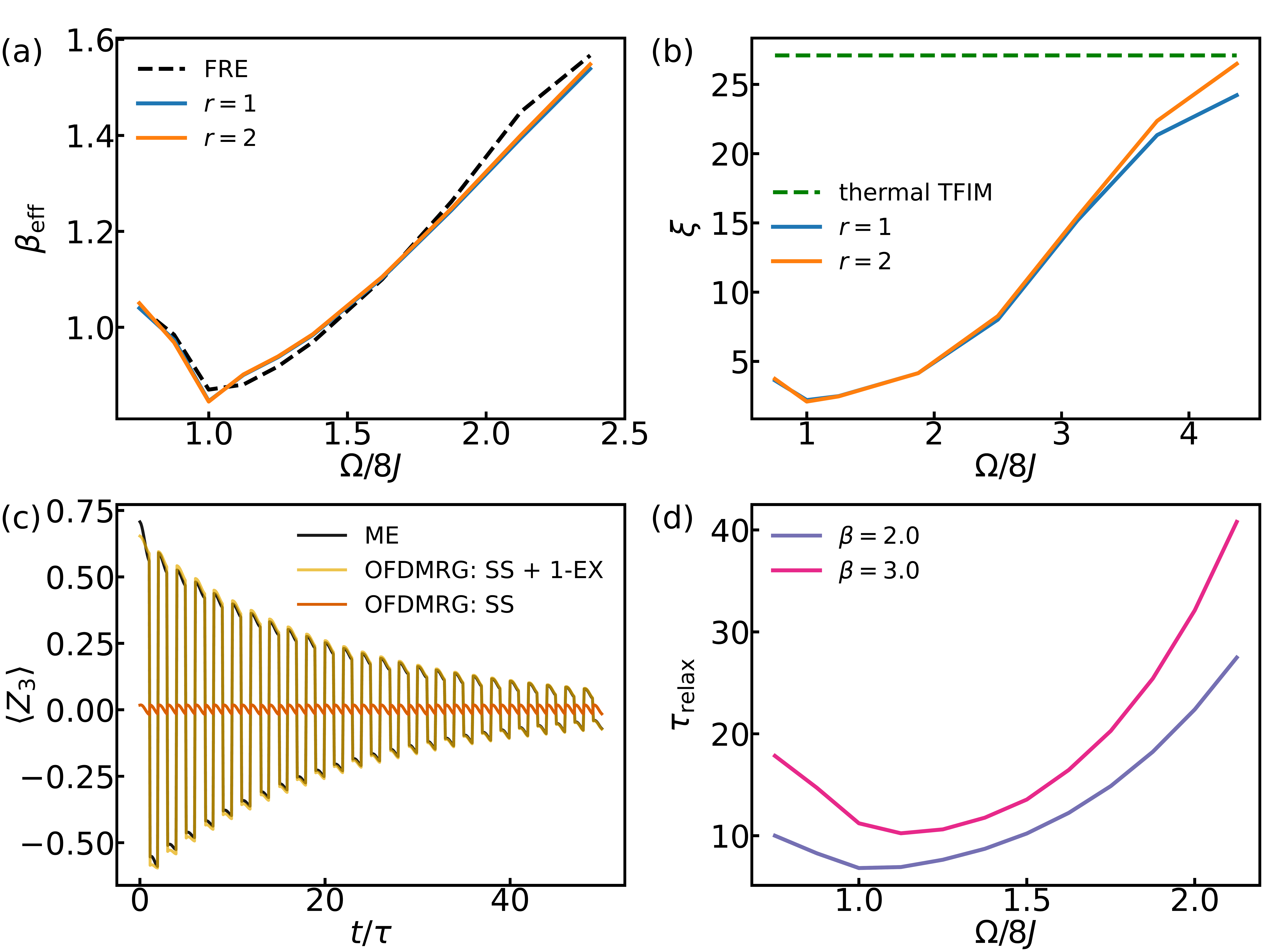}
	\caption{{\bf OFDMRG for dissipative DTC model} Eq.~\ref{eq:DTCHamiltonian} for $J=1$, $h=0.5$, $\omega_c=2$, and unless otherwise specified: $\beta=2$ and $r=2$.
	(a) Comparison between effective temperature $\beta_{\rm eff}$ of the dissipative DTC model calculated by OFDMRG method and that from solving FRE, with $L=11$, $g=0.05$, $\gamma=0.2$, 
	%$\(J, h, g, \gamma, \omega_c, \beta\)=\(1.0, 0.5, 0.05, 0.2, 2.0, 2.0\)$, 
	and $\(N_c, \chi\)=\(1, 16\)$. 
	(b) Correlation lengths $\xi$ of the dissipative DTC model for $L=31$, 
	%$\(J, h, g, \gamma, \omega_c, \beta\)=\(1.0, 0.5, 0.05, 0.2, 2.0, 2.0\)$, 
	$g=0.05$, $\gamma=0.2$,
	and $\(N_c, \chi\)=\(1, 8\)$. The correlation length for a thermal state of transverse-field Ising model (TFIM) with $\beta=2$ is given as a reference. 
	(c) Comparison between transient dynamics of $\langle Z_3\rangle$ calculated by OFDMRG method and by the exact evolution of master equation (ME) for $L=5$, $g=0.2$, $\gamma=2$, $\beta=5$, high-frequency ($\Omega=10$),
	%$\(J, h, g, \gamma, \omega_c, \beta, \Omega\)=\(1.0, 0.5, 0.2, 2.0, 2.0, 5.0, 10.0\)$, 
	and $\(N_c, \chi\)=\(2, 16\)$. 
	(d) Relaxation time of the dissipative DTC model for $L=21$, $g=0.2$, $\gamma=2$,
	%$\(J, h, g, \gamma, \omega_c\)=\(1.0, 0.5, 0.2, 2.0, 2.0\)$, 
	and $\(N_c, \chi\)=\(1, 16\)$.
	}
	\label{fig:XDTC}
	\vspace{-0.2in}
\end{figure}

\paragraph{Dissipatively-stabilizing a discrete time-crystal (DTC)}
Having benchmarked the performance of the OFDMRG approach, we now turn to the question of whether a pre-thermal dynamical phase can be stabilized by coupling to a cold bath. As an example, we study a model for a pre-thermal DTC model~\cite{Else2017} coupled to a thermal bath. For the system part, we consider one-dimensional Ising model driven by periodic $\pi$-pulses with generic perturbation breaking the $\mathbb{Z}_2$ symmetry, which serves as a prototypical model for the pre-thermal DTC~\cite{Else2017}
\begin{align}\label{eq:DTCHamiltonian}
	H(t) =& \sum_i\[\frac{\pi}{2}\sum_n\delta(t-n\tau)X_i-JZ_iZ_{i+1}+hZ_if(t)+gX_i\],
\end{align}
where $f(t)= \(1-\cos\Omega t\)$.  
Various disordered and/or long-range interacting incarnations of this Hamiltonian have been studied in previous theoretical studies and implemented experimentally in a variety of systems~\cite{Else2020,Harper2019} to study MBL and prethermal mechanisms for stabilizing DTC order in (approximately) closed systems.

Here, we introduce dissipation by coupling each spin, via coupling strength $\gamma$, to a separate ohmic bath with spectral function $J(\e)=\frac{\e}{\e_0}e^{-|\e|/\omega_c}/\(1-e^{-\beta\e}\)$, where $\beta=1/T$ is the inverse temperature of the bath, $\e_0$ is a characteristic energy scale, and $\omega_c$ is the local bandwidth of the bath, which will play an important role in controlling the steady state~\cite{Seetharam2015}.
%For the simplicity, we assume each spin in the chain is coupled an independent thermal bath which gives the jump operator given in Eq.(\ref{eq:jumpoperator}),  $\mathcal{X}_i=X_i$ and $g(t)$ corresponds to the spectral density with an ohmic form and an exponential cutoff 
%$J(\e)=\frac{\e}{\e_0}\frac{ e^{-\e/\omega_c}}{1-e^{-\beta\e}}$.
%where $\omega_c$ sets cutoff of the bath spectral and will be a crucial parameter for engineering the steady state~\cite{Seetharam2015}, and $\beta = 1/T$ is the inverse bath temperature.
We compute the effective time-dependent jump operators for this model using a Born-Markov approximation (see Appendix~\ref{appendix:microscopicjumpoperator} for details), and then truncate these to a finite range of $(2r+1)$ sites to incorporate into the OFDMRG procedure. 

The singular $\delta$-train has unbounded Fourier spectrum, which would be long range in frequency-space. However, for models with smooth $f(t)$ satisfying $f(0)=0$, we can cure this by transforming into a rotating frame of the $\delta$-function $X_\pi$-pulses. In the rotating frame the periodicity is doubled to $2\tau$, but there is a dynamical symmetry: $H(t+\tau) = XH(t)X$ with $X=\prod_i X_i$.
In the DTC phase~\cite{Else2020,Harper2019}, this dynamical symmetry is spontaneously broken, resulting in persistent period-doubled oscillations, and manifesting in long-range mutual information between distant spins~\cite{Else2016}. 
Unlike the long-range interacting pre-thermal DTC model realized recently in trapped-ion experiments~\cite{kyprianidis2021observation} such spontaneous symmetry breaking is forbidden in any short-range interacting $1d$ system that thermalizes to a finite temperature.
Instead, one expects the length- and time-scales for these signatures to diverge if the system is successfully cooled to a pre-thermal ground-state. 
The criterion of cooling near the pre-thermal ground is also required to realize dynamical Floquet topological phases (in any dimension), whose properties rely crucially on quantum coherence and entanglement.

Our goal is to assess whether and under what conditions the resulting NESS resembles a low-temperature Floquet Gibbs state with extended range DTC correlations. 
To this end, we compute (i) the NESS entropy $S_{\rm ss}=-{\rm Tr}\(\rho_{\rm ss}\log\rho_{\rm ss}\)$;
(ii) the NESS DTC spatial correlation length $\xi$ defined by fitting the averaged correlation function $\overline{\<Z_{j+x}Z_j\>}$ to the form $e^{-x/\xi}$ (shown in Fig.~\ref{fig:XDTC}(a)(b)). The NESS results are compared to properties of a thermal state  $\rho_\text{thermal} = \frac{1}{\mathcal{Z}}e^{-\beta D}$ where $D$ is the effective Floquet Hamiltonian obtained by performing a high-frequency (ven Vleck) expansion to the second order. 
We can extract an effective inverse temperature $\beta_{\rm eff} = 1/T_{\rm eff}$ by comparing the system entropy $S_{\rm ss}$ to the thermal entropy of $D$.
$D$ takes the form of a transverse-field Ising model with ordered ground-state, and the characteristic energy scale to make a local spin-flip excitation of $D$ is $4J$, which will result enhanced drive-induced heating when $\Omega/2 \approx 4J$, resulting in enhanced $T_{\rm eff}$.
We also compare results to solutions to an approximate Floquet rate equation (FRE)~\cite{Shirai2020,Ikeda2021,Mori2022} (for $L=11$) obtained from a Fermi-Golden rule treatment bath-induced transition rates between eigenstates of the effective system Hamiltonian $D$ in a rotating frame, which neglects off-diagonal coherences in the system density matrix (See details in Appendix \ref{appendix:FREandFGS}).
As driving frequency increases beyond $8J$, this heating is suppressed, and the system's $\beta_{\rm eff}$ asymptotes to that of the bath (note that, simulating colder temperatures requires keeping a larger spatial extent, $r$, to the ab initio computed jump operators), and $\xi$ increases towards the thermal correlation length of $\rho_\text{thermal}$ at the bath temperature. 
Importantly, the Floquet Gibbs state arises only when the local bath bandwidth satisfies $\omega_c\ll\left\vert\frac{\Omega}{2}-4J\right\vert$, so that bath-assisted drive-induced heat absorption processes are suppressed (See details in Appendix \ref{appendix:FREandFGS}).

We further explore the long-time DTC dynamics, through asymptotic decay rate $\tau_{\rm relax} = -\({\rm Re}\lambda_1\)^{-1}$ of period doubled oscillations obtained by computing the first excited eigenstate $|\rho_1\rrangle$, as well as the explicit dynamics of $\<Z_j(t)\>$ for the $|\rho(t)\rrangle = |\rho_{\rm ss}\rrangle + e^{-\lambda_1 t}\llangle \rho_I|\rho_1\rrangle |\rho_1\rrangle$, which captures the long-time dynamics from an initial product state: 
%$\rho_I = |\psi_I\>\<\psi_I|$ with $|\psi_I\> = \otimes_i(\sin\frac{\pi}{8}|\up\>+\cos(\frac{\pi}{8})|\down\>)$.
$\rho_I = \prod_i (\sin\frac{\pi}{8}|\up\>+\cos(\frac{\pi}{8})|\down\>)(\sin\frac{\pi}{8}\<\up|+\cos(\frac{\pi}{8})\<\down|)$.
As shown in Fig.~\ref{fig:XDTC}(c), the dynamical results are compared against exact master equation simulations (for $L=5$, close to the limit of a single workstation).
We observe quantitative agreement between the time-dependent dynamics of the excited-state OFDMRG method with the master equation simulations, confirming that the long-time dynamics is indeed dominated by the first decaying mode. 
Further, in Fig.~\ref{fig:XDTC}(d), we observe that the DTC time scale increases with driving frequency $\Omega$ (for $\Omega/2>4J$), asymptoting to a finite time-scale that increases as the bath is cooled.

\paragraph{Discussion and outlook} 
These results confirm the expectation that there is a parameter regime of large driving frequency ($\Omega \gg 8J$), moderate bath bandwidth ($\omega_c\ll\left\vert\frac{\Omega}{2}-4J\right\vert$), and moderately-weak system bath coupling ($e^{-J/\Omega}\ll \gamma \ll J$) where coupling the pre-thermal DTC model to a bath successfully produces a Floquet Gibbs-like state with temperature close to that of the bath (See details in Appendix \ref{appendix:FREandFGS} and \ref{appendix:highfrequencyexpansionfortotalH}). 
Further, the OFDMRG method successfully captures this behavior in system sizes that greatly exceed those accessible by exact master equation simulations (here we simulated up to $L=31$ on a single computer, which would be limited to $L\lesssim 6$ for exact computation). 

We expect this technique to be useful in designing realistic realizations of dissipatively-stabilized dynamical phases in solid-state devices and atomic physics quantum simulators.
The OFDMRG also permits a controlled means to assess the validity of various approximation methods such as Floquet rate equations which could potentially be used beyond $1d$.
Natural future targets for extending the OFDMRG method include studying NESS of quasiperiodically driven systems~\cite{Dumitrescu2018,Else2020b, Friedman2022} (with multiple frequency-space directions), and incorporating non-Markovian effects~\cite{Mizuta2021,Schnell2021}.

\vspace{4pt}\noindent {\it Acknowledgments} We thank Brayden Ware and Romain Vasseur for insightful discussions. This work was supported by NSF DMR-1653007 and the Alfred P. Sloan Foundation through a Sloan Research Fellowship (A.C.P.).

\bibliography{Open-Floquet}

\appendix
\onecolumngrid

\section{Floquet theorem for open quantum systems}
\label{appendix:floquettheorem}
Consider a time-periodic Liouvillian superoperator
\begin{align}
	\frac{d\rho}{dt}=\mathcal{L}(t)[\rho].
\end{align}
By integrating both sides, one can define a  time evolution superoperator 
\begin{align}
	\mathcal{E}(t,t_0)=\mathcal{T}\exp\(\int_{t_0}^t ds \mathcal{L}(s)\),
\end{align}
with
\begin{align}
	\rho(t)=\mathcal{E}(t, t_0)[\rho(t_0)].
\end{align}
%One  quick  property  from  the  time-periodic  Liouvillian  is  that, since $\mathcal{E}(t+\tau, t_0+\tau)=\mathcal{E}(t, t_0)$, only in-period  time-evolution $t\in\[t_0, t_0+\tau\]$ matters.  
For general initial and final time, one can always decompose the evolution as multiple one-period evolution and some in-period evolution
\begin{align}
	\mathcal{E}(t_f, t_0)=\mathcal{E}(t+n\tau, t_0)=\mathcal{E}(t, t_0)\mathcal{E}(t_0+\tau, t_0)^n,
\end{align}
where $t\in\[t_0, t_0+\tau\]$. Here the one-period evoluion superopertor $\mathcal{E}(t_0+\tau, t_0)$ is of particular interest. 
 
One can obtain a complete basis $\{\rho_\alpha(t_0)\}$ of density matrix by diagonalizing the one-period evolution superoperator at $t_0$
	\begin{align}
		\mathcal{E}(t_0+\tau, t_0)[\rho_\alpha(t_0)]=e^{-\lambda_\alpha \tau}\rho_\alpha(t_0).
	\end{align}
The eigenstates at arbitrary time $t$ with the same spectrum are given by $\rho_\alpha(t)=\mathcal{E}(t,t_0)\rho_\alpha(t_0)$:
\begin{align}
	\mathcal{E}(t+\tau,t)[\rho_\alpha(t)]=\mathcal{E}(t,t_0)\mathcal{E}(t_0+\tau,t_0)\mathcal{E}(t_0,t)[\rho_\alpha(t)]=\mathcal{E}(t,t_0)\mathcal{E}(t_0+\tau,t_0)[\rho_\alpha(t_0)]=e^{-\lambda_\alpha \tau}\rho_\alpha(t).
\end{align}

Analogous to the Bloch state, each eigenstate satisfying $	\rho_\alpha(t+\tau)=e^{-\lambda_\alpha \tau}\rho_\alpha(t)$ can be decomposed as
\begin{align}
	\rho_\alpha(t)=e^{-\lambda_\alpha t}\tilde{\rho}_\alpha(t),
\end{align}
where $\tilde{\rho}_\alpha(t)=\tilde{\rho}_\alpha(t+\tau)$. Importantly, one can thus expand $\tilde{\rho}_\alpha(t)$ into Fouerier series
\begin{align}
	\tilde{\rho}_\alpha(t)=\sum_ne^{in\Omega t}\rho^n_\alpha,
\end{align}
which allows one to solve the time-periodic Liouvillian as a time-independent problem in an extended Hilbert space $\mathcal{H}^2\times \mathbb{Z}$ with  basis 
$\rho^n_\alpha$.

\section{OFDMRG Implementation Details: Preconditioning heuristics and convergence checks}\label{appendix:warmupconvergence}
Besides adding penalizing terms to meet the trace constraints, in the implementation, we also use some warm-up steps to avoid getting stuck in local minima: (i) solve $N_c=0$ first, i.e. time-averged Lindbladian as a initial guess and then gradually increase $N_c$ until get converged result; (ii) gradually increase the bond dimension of the MPO to obtain good convergence; (iii) For the weak dissipation case, i.e. $\gamma\sim\vert L_\alpha\vert^2$ is small, the dissipative gap is usually even smaller than the spectral gap of Hamiltonian, and then it is easy to become trapped by a local minima. To overcome this, we start with a large $\gamma\sim1$ and decrease it gradually.

As mentioned in the main text, the numerical error of our method arises from cut-offs in the frequency-space and the MPO bond dimension. The error from the frequency-space trucation is controlled by $\vert \rho^{N_c}\vert/\vert \rho^0\vert$. In Fig. \ref{fig:converge}(a) we can see  $\vert \rho^n\vert/\vert \rho^0\vert$ decays exponentially in $|n|$ away from $n=0$, indicating localization in frequency-space compatible with truncation to finite $|n|<N_c$. 
Fig. \ref{fig:converge}(b) shows the exponential decaying of Schmidt components in the MPO ansatz, which allows the large system size calculation. In practice, we set $10^{-3}$ as a tolerance bar for determining if the calculation is converged.

\section{Positivity of density matrices}\label{appendix:positivity}
\begin{figure}[htp]
	\includegraphics[width=\columnwidth]{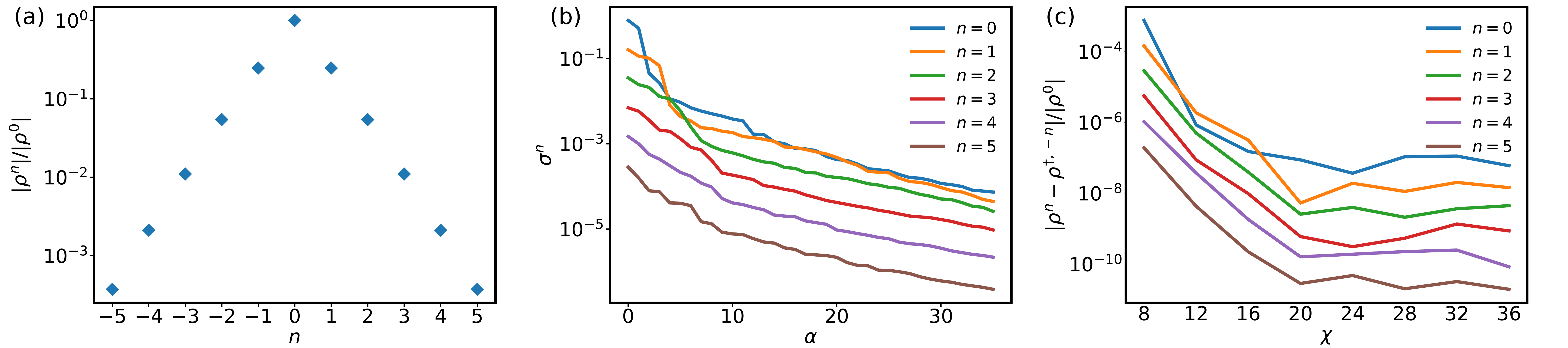}
	\caption{{\bf Convergence and positivity of density matrices -- } (a) The relative weight $\frac{\vert\rho^n\vert}{\vert\rho^0\vert}$ for each fourier component $n$ of Floquet density matrix for $N_c=5$ and $\chi=36$. (b) Schimt components $\sigma^n$ for each fourier component $n$ of Floquet density matrix for $N_c=5$ and $\chi=36$. (c) Non-Hermicity $\frac{\vert\rho^n-\rho^{\dagger,-n}\vert}{\vert\rho^0\vert}$ for each fourier component $n$ of Floquet density matrix versus $\chi$. }
	\label{fig:converge}
\end{figure}

Besides having unit trace, a physical density matrix must also satisfy a positivity condition, which ensures the positive occupation on each eigenstate, this condition is generally NP-hard (in system size) problem to even check. In our algorithm, the positivity of density matrix is not manifestly guaranteed (compared to the local purified tensor network ansatz~\cite{Werner2016}). However, as the extended Lindbladian is a complete positive superoperator, the exact steady-state solution is a positive fixed point and thus if our procedure does not become stuck in any local minima, and if there is no other dark state solution, we expect the solved density matrix is positive at least up to the numerical error from frequency-space and bond-dimension truncation.
Although the direct check on the positivity is NP-hard, we provide an indirect check on the non-Hermiticity $\vert\rho^n-\rho^{\dagger,-n}\vert/\vert\rho^0\vert$ as shown in Fig.~\ref{fig:converge}(c). We observe the non-Hermiticity of density matrix is small and consistent with error induced by truncations on the frequency-space and bond dimension, which is at the same order of $\sigma^n(\chi)\vert \rho^{n}\vert/\vert \rho^0\vert$. Besides, all the physical observables we demonstrated in the main text take physical values with tolerance set by truncation errors.

\section{General jump operators from microscopic model and MPO construction for dissipative discrete time-crystal model}
\label{appendix:microscopicjumpoperator}

In the driven-dissipative Ising model, we studied a phenomenological Lindbladian. However, to realistically capture the interactions between a system and a thermal bath at a specific temperature, one must microscopically derive the resulting master equation. The most commonly presented derivations of the Lindblad master equations~\cite{Breuer2002} are usually based on two main approximations: (i) the Born-Markov approximation (BMA), which requires the bath retains in approximate equilibrium while the system is dissipating, that is $\tau_R\gg\tau_B$
where $\tau_R\sim\gamma^{-1}$ is 
the relaxation time from system to bath and $\tau_B$ is the bath correlation time; and (ii) the rotating wave approximation (RWA) requiring the intrinsic time scale of the system $\tau_X\sim\vert\e_i-\e_j\vert^{-1}$ is small compared to the relaxation time, so that the oscillation between system energy levels can be smeared out within the relaxation process.
The BMA is valid for weak system-bath coupling and Markovian baths (the latter condition is independent of the system). However, the RWA is usually broken down in quantum many-body systems, where the level spacing goes to infinite small in the thermodynamic limit. Recently, some important progresses have been made to derive the Lindblad master equation without the RWA and its error bound is consistent with the BMA only\cite{Nathan2020, Mozgunov2020}.  Here we leverage the Lindblad equation derivation from Ref~\cite{Nathan2020} to describe the Floquet-dissipative system of interest. 

Consider a system-plus-bath Hamiltonian $H(t)=H_S(t)+H_B+H_{SB}$, where $H_S(t)$ is the Hamiltonian for the system, including periodic driving, $H_B$ is for the thermal bath, and $H_{SB}$ represents the system-bath interactions, which can be generally decomposed as
\begin{align}
	H_{SB}=\sum_i\sqrt{\gamma} \mathcal{X}_i\otimes B_i,
\end{align}
where $\gamma$ is a parameter representing the strength of system-bath coupling, operator $\mathcal{X}_i$ acts only on the system, and operator $B_i$ acts only on the bath. Note that the behavior of bath is characterized completely by the bath correlation function $J_{ij}(t-s)=\langle B_i(t)B_j(s)\rangle$ and the corresponding spectral desity $J_{ij}(\omega)=\int dt J_{ij}(t)e^{i\omega t}$. As shown in Ref.\cite{Nathan2020}, a universal Lindblad master equation can be derived up to the accuracy of the BMA, i.e. $\mathcal{O}(\gamma^2\tau_B)$:
\begin{align}
	\mathcal{L}(t)[\rho]=&-i\[H(t)+\Lambda(t),\rho\]+\sum_i\(L_i(t)\rho L_i^\dagger(t)-\frac{1}{2}\left\{L^\dagger_i(t) L_i(t), \rho\right\}\),
\end{align}
with the Lamb shift $\Lambda(t)$ and jump operators given by:
\begin{align}
	\Lambda(t) =\sum_i \frac{\gamma}{2i}\int_{-\infty}^\infty ds ds'U(t,s)\mathcal{X}_iU(s,s')\mathcal{X}_iU(s',t)\phi(s-t, s'-t),
\end{align}
\begin{align}\label{eq:jumpoperator}
	L_i(t) = \sqrt{\gamma}\int_{-\infty}^\infty ds g(s-t)U(t,s)\mathcal{X}_iU(s,t),
\end{align}
where $U(t,s)=\mathcal{T}\exp\(-i\int_s^tdt'{H}(t')\)$ is the time evolution operator for the system, $\phi(t,s) = g(t)g(-s){\rm sgn}(t-s)$ and $g(s)$ is the jump correlator given by the Fourier transformation of the squared root of the bath spectral density $J(\omega)$
\begin{align}
	g(t) = \frac{1}{2\pi}\int_{-\infty}^\infty d\omega g(\omega)e^{-i\omega t}, \quad g(\omega)=\sqrt{J(\omega)},
\end{align}
Note that in the weak coupling limit, the Lamb shift does not significantly change the NESS, thus we neglect the Lamb shift in our discussion, however $\Lambda(t)$ could be straightforwardly incorporated into $H(t)$ if desired.

For specificity, we consider the dissipative DTC mentioned in the main text Eq.~\ref{eq:DTCHamiltonian}, coupled to independent thermal baths by $\sqrt{\gamma} X_i\otimes B_i$, and the bath spectral function is of Ohmic form $J(\e)=\frac{\e}{\e_0}e^{-|\e|/\omega_c}/(1-e^{-\beta\e})$. In the rotating frame of the $X_{\pi}$-pulses, the DTC Hamiltonian (Eq.~\ref{eq:DTCHamiltonian}) and jump operators for the universal Lindblad equation read:
\begin{align}\label{eq:rotatingHamiltonian}
	\tilde{H}(t)=\sum_i\[-JZ_iZ_{i+1}+gX_i+hZ_if(t)(-1)^{1+[t/T]}\],
\end{align}
and
\begin{align}\label{eq:rotatingjumpoperator}
	\tilde{L}_i(t)=\sqrt{\gamma}\int_{-\infty}^\infty ds g(s-t)\tilde{U}(t,s){X}_i\tilde{U}(s,t),
\end{align}
where $\tilde{U}(t,s)=\mathcal{T}\exp\(-i\int_s^tdt'\tilde{H}(t')\)$ is time-evolution operator for the Hamitonian in the rotating frame. 

In general, jump operators given in Eq.(\ref{eq:rotatingjumpoperator}) are not strictly local, but however, have exponential decaying tails out to larger distances. 
%A standard Lieb-Robinson bound provides a controllable approximation to truncating these exponential tails beyond a finite range that depends on the detailed parameters.
%A natural local truncation of jump operators is given by only allowing evolution in the region
In our implementation we truncate these tails outside a region, $\[i-r, i+r\]$:
\begin{align}\label{eq:jumpoperatorreduced}
	\tilde{L}_{i,r}(t)=\sqrt{\gamma}\int_{-\infty}^\infty ds g(s-t)\tilde{U}_r(t,s){X}_i\tilde{U}_r(s,t),
\end{align}
with $\tilde{U}_r(t,s)=\mathcal{T}\exp\(-i\int_s^tdt'\tilde{H}_r(t')\)$ and $\tilde{H}_r$ is only defined in $\[i-r, i+r\]$. An error bound for such a local truncation on the jump operator is given by considering the Lieb-Robinson bound from cutting the exponential tail outsides the light cone of system time-evolution (See Appendix \ref{appendix:errorbound} for the derivation):
\begin{align}
	\vert\vert \tilde{L}_i(t)-\tilde{L}_{i,r}(t)	\vert\vert\lesssim Ce^{-\kappa(r-v\tau_B)},
\end{align}
where $C$ is a non-universal constant, $\tau_B$ is the bath correlation time, and $v$ and $\kappa$ depend only on the system Hamiltonian.  For the DTC model, $v\sim2J$ and $\kappa\sim\log\vert J/g\vert$, the truncation is valid when $r\gg 1/\log\vert J/g\vert+2J\tau_B$, which requires small transverse field and bath correlation time. On the other hand, since the bond dimension of a generic MPO representation for superoperators grows exponentially with the active range as $ \chi_O=16^{r}$, this translates to a exponential growth of computational complexity with the bath correlation time $\tau_B$. In practice we are unable to perform computations with $r\geqslant3$, and consider only $r=1$ and $r=2$ cases in our implementation. Yet, this range is sufficient to accurately capture the behavior of moderately-low temperature baths ($\beta\lesssim\frac{\pi}{J}$, as lower temperature results in long bath correaltions) and for moderately small transverse fields (as large transverse fields result in large $v$).
%But even for $r=1$ and $r=2$ cases, the Lieb-Robinson bound is still valid except for large transverse field and low temperature ($\beta\gtrsim\frac{\pi}{J}$) which is also ruled out by the break of Markovian approximation.

%We are interested not only in the stroboscopic behavior at fixed times within the period, but also in the full micromotion of the operators -- as this micromotion can contain important information about dynamical phases and topology.
Since we are particularly interested in the high-frequency regime (possibly in an appropriately rotating frame), where driving-induced heating is much larger than the relaxation time from the system-bath coupling $\sim\gamma^{-1}$, we adopt a high-frequency expansion to simplify 
%Since we require not only the stroboscopic behavior but also full micromotion within the period (at least the truncated Hilbert space is required, it is still a giving nontrivial task to give the explicit expression for jump operators after the truncation Eq.(\ref{eq:jumpoperatorreduced}). Considering we are particularly interested in the high-frequency regime where driving-induced heating is much larger than the relaxation time from the system-bath coupling $\sim\gamma^{-1}$, the high-frequency expansion can be adopted to simplify 
Eq.(\ref{eq:jumpoperatorreduced}) as:
\begin{align}
	\tilde{U}_{r}(t,s)\approx e^{-iK_r(t)}e^{-iD_r(t-s)}e^{iK_r(s)},
\end{align}
where $D_r$ is the time-independent effective Hamiltonian and the $K_r(t)$ is the periodic kicking operator. (In our implementation we keep the expansion upto $\mathcal{O}(\Omega^{-2})$ ).  A significant advantage of the high-frequency expansion is that we obtain a spectral representation of the jump operators with respect to eigenbasis of the time-independent Floquet Hamiltonian $D_r$. 
%Consider the frame transformation by the kicking operator $e^{-iK_r(t)}$ and then we can 
Define the transformed jump operator $M_{i, r}(t)$ in the interacting frame or the kicking operator $e^{-iK_r(t)}$  as
\begin{align}\label{eq:jumpoperatorhighfrequency}
	M_{i, r}(t) &\approx e^{iK_r(t)}\tilde{L}_{i,r}(t)e^{-iK_r(t)}\nonumber\\
	&=\sqrt{\gamma}\int_{-\infty}^\infty ds g(s-t)e^{-iD_r(t-s)}\tilde{\mathcal{Y}}_i(s)e^{iD_r(t-s)}\nonumber\\
	&=\sqrt{\gamma}\sum_k e^{ik\omega t/2}\sum_{mn}g\(\tilde{\e}_n-\tilde{\e}_m-\frac{k\Omega}{2}\)\tilde{\mathcal{Y}}_{i, mn}\vert m\rangle\langle n\vert,
\end{align}
where $\tilde{\e}_n$ is the spectrum of $D_r$ and $\tilde{\mathcal{Y}}_i(s) = e^{iK_r(s)}X_ie^{-iK_r(s)}$. Then the approximated jump operator in the rotating frame of the $X_\pi$-pulse, $\tilde{L}_{i, r}(t)\approx e^{-iK_r(t)}M_{i,r}(t)e^{iK_r(t)}$ ,from Eq.(\ref{eq:jumpoperatorhighfrequency}) gives a strictly-local operator which is compatible with our MPO-based algorithm.

\subsection{Error bound for the local truncation on jump operators }\label{appendix:errorbound}
We next bound the error due to truncating the jump operators to a finite spatial region.
Recall that the general jump operator is given by a convolution between jump correlation function $g(t)$ and system-bath interaction operator $\mathcal{X}_i$
\begin{align}\label{seq:jumpoperator}
	{L}_i(t)=\sqrt{\gamma}\int_{-\infty}^\infty ds g(s-t){U}(t,s){\mathcal{X}}_i(s){U}(s,t),\quad  U(t,s)=\mathcal{T}\exp\(-i\int_s^tdt' H(t')\),
\end{align}
whose effective range is determined by the bath correlation time encoded in $g(t)$ and the Lieb-Robinson bound of the system. 
%In this section, we give an derivation of the error bound for setting a local truncation on $L_i(t)$.

For simplicity, we assume the system-bath interaction operator $\mathcal{X}_i$ only acts on $i$ site and the system Hamiltonian is nearest-neighbor interacting, $H(t) =\sum_jh_{j,j+1}(t)$. Then we follow the procedure in Ref.\cite{Gong2022} by considering a modified Hamiltonian with two cut bonds $H_{r}=H-h_{i-r-1,i-r}-h_{i+r, i+r+1}$. Since $H$ is nearest-neighbor interacting, the support of operators in $[i-r,i+r]$ evolved by $H_r$ is confined in the region $\[i-r, i+r\]$, which gives a local truncation on jump operators. The corresponding error from replacing $H$ with $H_r$ is given by
\begin{align}\label{seq:errorbound1}
	\vert\vert L_i(t)-L_{i,r}(t)\vert\vert\leqslant &\sqrt{\gamma}\int_{-\infty}^\infty ds g(s-t)\vert\vert{U}(t,s){\mathcal{X}}_i(s){U}(s,t)-{U}_r(t,s){\mathcal{X}}_i(s){U}_r(s,t) \vert\vert \nonumber\\
	=&\sqrt{\gamma}\int_{-\infty}^\infty ds g(s-t) \left\vert\left\vert \[U_r^\dagger(t,s)U(t,s), {\mathcal{X}}_i(s)\] \right\vert\right\vert\nonumber\\
	\leqslant&\sqrt{\gamma}\int_{-\infty}^\infty ds g(s-t) \int_{s}^tdt'\left\vert\left\vert\[V_s(t'), {\mathcal{X}}_i(s) \]\right\vert\right\vert,
\end{align}
where
\begin{align}
	V_s(t) = U^\dagger(t,s)\(h_{i-r-1,i-r}+h_{i+r, i+r+1}\)U(t,s),
\end{align}
is the time-dependent generator of $U^\dagger(t,s)U_r(t,s)$. In the last step, we have invoked the Kubo identity:
\begin{align}
	\[A, e^{\beta H}\] = \int^\beta_0 d\lambda e^{(\beta-\lambda)H}\[A, H\]e^{\lambda H}.
\end{align}
The general form of the Lieb-Robinson bound is given by
\begin{align}
	\vert\vert\[\mathcal{O}_X(t),\mathcal{O}_Y  \]\vert\vert\leqslant C_0 e^{-\kappa S_{XY}}\(e^{\kappa v\vert t\vert/}-1\),
\end{align}
where $S_{XY}$ is minimum distance between operator $\mathcal{O}_X$ and $\mathcal{O}_Y$, and $\xi$ and $v$ are constants  that only depend the system Hamiltonian. Applying the Lieb-Robinson bound to Eq.(\ref{seq:errorbound1}), we obtain:
\begin{align}
	\vert\vert L_i(t)-L_{i,r}(t)\vert\vert\leqslant&\sqrt{\gamma}\int_{-\infty}^\infty ds g(s-t) \int_{s}^tdt'C_0 e^{-\kappa S_{XY}}\(e^{\kappa v\vert t'-s\vert}-1\)
	%\nonumber\\
	\lesssim C e^{-\kappa\(r-v\tau_B\)},
\end{align}
which gives an error bound for replacing $L_i(t)$ by $L_{i,r}(t)$.

\section{Floquet rate equation and conditions for Floquet-Gibbs states}
\label{appendix:FREandFGS}
When the system-bath coupling $\gamma$ is significantly smaller than the system characteristic local energy $J$,  a Floquet rate equation (FRE)  can be derived from the Floquet Lindblad equation perturbatively\cite{Mori2022,Shirai2020} , by neglecting off-diagonal coherence between Floquet eigenstates (of the system Hamiltonian), and treating only incoherent transitions between different diagonal entries of the density matrix, $\rho_{qq}$ where $\e_q$ is a fixed quasi-energy of the system Hamiltonian with index $q$.
This FRE can be used to obtain intuition about various heating and cooling processes and their associated rates to identify a regime where the system can be effectively cooled into its pre-thermal ground-state.

The steady-state condition for the FRE is:
\begin{align}
	\frac{d\rho_{qq}}{dt} = 0=\sum_{p}\rho_{pp}R_{pq}-\rho_{qq}\sum_{p}R_{qp},
\end{align}
where the Fermi-Golden rule transition rates are:
\begin{align}
	R_{pq}= \sum_{i, k}J\(\e_q-\e_p- k \Omega\)\vert \langle p\vert\mathcal{X}^k_{i}\vert q\rangle\vert^2.
\end{align}
where $\e_p$ and $\e_q$ are quasi-energies of the system Floquet Hamiltonian. Since we are particularly interested in the weak dissipation regime where the BMA holds, it is natural to expect the FLE gives similar results to the FRE.
Importantly, the OFDMRG approach can go beyond the FRE, and can also be used to assess the validity of the approximations made in the simpler FRE. In a complementary way, the FRE can be used to gain intuition about the asymptotic behavior observed in the OFDMRG approach.

Specifically, for the dissipative DTC model, we consider the FRE in the $X_\pi$-pulse rotating frame and the high-frequency kicking frame:
\begin{align}\label{eq:FloquetrateequationI}
	0=\tilde{\rho}_{qq}\sum_{p}\tilde{R}_{qp}-\sum_{p}\tilde{\rho}_{pp}\tilde{R}_{pq},
\end{align}
\begin{align}\label{eq:FloquetrateequationII}
	\tilde{R}_{pq}= \sum_{i, k}J\(\tilde{\e}_q-\tilde{\e}_p- \frac{k}{2} \Omega\)\vert \langle p\vert\tilde{\mathcal{Y}}^k_{i}\vert q\rangle\vert^2.
\end{align}
where $\tilde{\e}_p$ and $\tilde{\e}_q$ are replaced by eigenenergies of the effective Hamiltonian $D$. In practice, we solve Eq.($\ref{eq:FloquetrateequationI}$) and Eq.($\ref{eq:FloquetrateequationII}$) by the exact diagonalization. 

For the static model with $\Omega=0$, the Fermi-golden-rule like transition rate $\tilde{R}_{pq}$ satisfy the detailed balance condition
\begin{align}\label{eq:detailedbalance}
	\frac{\tilde{R}_{pq}}{\tilde{R}_{qp}}=e^{-\beta\(\tilde{\e}_q-\tilde{\e}_p\)},
\end{align}
with which the rate equation gives to the Gibbs density matrix solution
\begin{align}
	P_p\equiv\rho_{pp}=\frac{e^{-\beta\tilde{\e}_p}}{Z},\quad Z\equiv\sum_pe^{-\beta\tilde{\e}_p}.
\end{align}
For the driven model with $\Omega>0$, there is driving quanta exchange process given by $J\(\tilde{\e}_q-\tilde{\e}_p-\frac{k}{2}\Omega\)$ terms,
which breaks the detailed balance in Eq.~\ref{eq:detailedbalance} and leads to deviations from Gibbs-like solutions for the steady state. When these processes are appreciable, the system steady state typically heats up to a higher effective temperature than the temperature of bath to which it couples. Thus, suppressing the bath-assisted heat absorption requires a small bath spectrum cutoff satisfying\cite{Shirai2016,Ikeda2021,Mori2022}
\begin{align}
	J\(\Lambda-\frac{k}{2}\Omega\)\sim \exp\(-\frac{1}{\omega_c}\left\vert\Lambda-\frac{k}{2}\Omega\right\vert\)\ll J\(\Lambda\) ,\quad\text{for }k\neq0,
\end{align}
which is equivalent to
\begin{align}\label{eq:conditionforFGS}
	\left\vert\Lambda-\frac{k\Omega}{2}\right\vert\gg \omega_c,\quad\text{for }k\neq0,
\end{align}
where $\Lambda$ is the local energy scale of system, which $\sim4J$ for the dissipative DTC model. Hence we can conclude that, to realize an approximate Floquet-Gibbs state, the driving frequency $\Omega$ needs not only to be larger than the system's energy scale $\Lambda$ but also must greatly exceed the local bandwidth of bath excitations, $\omega_c$. As one increases the driving frequency, the NESS becomes closer to a Floquet-Gibbs state with the bath temperature (as shown in the main text figures).
%To realize a specific steady state by balancing the driving-induced heating by a cold bath, one might expect the Floquet steady state features a Floquet-Gibbs state (FGS) for $D$ with the same temperature $1/\beta$ of the cold bath. However, in general the real steady state is not precisely a FGS, due to the driving-resonant term $J\(\tilde{\e}_q-\tilde{\e}_p- \frac{k}{2} \omega\)$ ($k\neq0$) contributing a recombination process between different Floquet sectors, which breaks the detailed balance between system and bath. Hence to obtain a FGS,  the driving-resonant term is required to be suppressed by the exponential cutoff of bath spectrum\cite{Shirai2016,Ikeda2021,Mori2022}, i.e. $\vert\tilde{\e}_q-\tilde{\e}_p- \frac{k}{2} \omega\vert\gg\omega_c$ for $k\neq0$, and as tuning the driving frequency, there is a crossover between non-FGS and FGS.

\section{High-frequency expansion for the system-plus-bath Hamiltonian}\label{appendix:highfrequencyexpansionfortotalH}
In this section, we provide another route to viewing the condition for Floquet-Gibbs states, by applying the high-frequency expansion to the system-plus-bath Hamiltonian and assuming driving frequency $\Omega$ is not only larger than the system's energy scale $\Lambda$ but also the bath spectrum cutoff $\omega_c$, i.e. $\Omega\gg\Lambda$ and $\Omega\gg\omega_c$.

Consider the generic Hamiltonian consisting of system and bath part with periodic external driving
\begin{align}
	H(t)=V(t)+\sqrt{\gamma}X_\alpha (t)B_\alpha+H_B,
\end{align}
where $V(t)$ is the time-periodic Hamiltonian for the system part, $X_\alpha$ and $B_\alpha$ are system-bath interaction operators acting in system and bath Hilbert space respectively, and $H_B$ is the bath Hamiltonian. 
Here, we do not consider the case where the bath is also driven.
In the high-frequency limit, the total evolution can be approximated by the evolution of an effective time-independent Hamiltonian after transforming into a periodically ``kicked" frame~\cite{Else2017,Mizuta2019} within a pre-thermal time $t_\ast= e^{-\mathcal{O}\(\Omega/{\rm max}\{\omega_c, \Lambda\}\)}$
\begin{align}
	U(t,s) \approx e^{-iK(t)}e^{-iD(t-s)}e^{iK(s)},
\end{align}
where $D$ is the effective Hamiltonian and $K(t)$ is the periodic kicking operator. 
The effective Hamiltonian and kicking operator is given by the van Vleck expansion
\begin{align}
 D =& {H}^0+\sum_{k\neq0}\frac{\[{H}^k, {H}^{-k}\]}{2k\Omega}+\sum_{k\neq0}\frac{\[\[{H^k},{H}^0\],{H}^{-k}\]}{2k^2\Omega^2}+\sum_{k\neq0}\sum_{q\neq k,0}\frac{\[\[{H^k},{H}^{q-k}\],{H}^{-q}\]}{3qk\Omega^2}+\mathcal{O}(\Omega^{-3}),
\end{align}
\begin{align}
	iK(t) =&\sum_{k\neq0}\frac{{H}^ke^{ik\Omega t/2}}{k\Omega} +\sum_{k\neq0}\frac{\[{H}^k, {H}^0\]e^{ik\Omega t/2}}{k^2\Omega^2}+\sum_{k\neq0}\sum_{q\neq k,0}\frac{\[{H}^q, {H}^{k-q}\]e^{i{k}\Omega t/2}}{2kq\Omega^2}+\mathcal{O}(\Omega^{-3}).
\end{align}
To obtain the effective description of the system degree of freedom which is what we really are interested in,  we only keep the first order of $\sqrt{\gamma}$ which is consistent with the Born-Markov approximation we will further apply
\begin{align}
	D =& D_V+H_B+\sqrt{\gamma}{X}^0B+\sqrt{\gamma}\sum_{k\neq0}\frac{\[{X}^k, {V}^{-k}\]+\[{V}^k, {X}^{-k}\]}{2k\Omega}B\nonumber\\&
	+\sqrt{\gamma}\sum_{k\neq0}\frac{\[\[{V^k},{X}^0\],{V}^{-k}\]+\[\[{X^k},{V}^0\],{V}^{-k}\]+\[\[{V^k},{V}^0\],{X}^{-k}\]}{2k^2\Omega^2}B\nonumber\\&
	+\sqrt{\gamma}\sum_{k\neq0}\frac{\[{X^k},{V}^{-k}\]}{2k^2\Omega}\frac{\[B,H_B\]}{\Omega}\nonumber\\&
	+\sqrt{\gamma}\sum_{k\neq0}\sum_{q\neq k,0}\frac{\[\[{X^k},{V}^{q-k}\],{V}^{-q}\]+\[\[{V^k},{X}^{q-k}\],{V}^{-q}\]+\[\[{V^k},{V}^{q-k}\],{X}^{-q}\]}{3qk\Omega^2}B\nonumber\\
	&+\mathcal{O}(\Omega^{-3},\gamma),
\end{align}
and
\begin{align}
	iK(t) =&iK_V(t)+\sqrt{\gamma}\sum_{k\neq0}\frac{{X}^ke^{ik\Omega t/2}}{k\Omega} B\nonumber\\
	&+\sqrt{\gamma}\sum_{k\neq0}\frac{\[{V}^k, {X}^0\]+\[{X}^k, {V}^0\]}{k^2\Omega^2}e^{ik\Omega t/2}B+\sqrt{\gamma}\sum_{k\neq0}\frac{{X}^k}{k^2\Omega}e^{ik\Omega t/2}\frac{\[B,H_B\]}{\Omega}\nonumber\\
	&+\sqrt{\gamma}\sum_{k\neq0}\sum_{q\neq k,0}\frac{\[{X}^q, {V}^{k-q}\]+\[{V}^q, {X}^{k-q}\]}{2kq\Omega^2}e^{i{k}\Omega t/2}B+\mathcal{O}(\Omega^{-3},\gamma),
\end{align}
where $D_V$ and $K_V$ are effective Hamiltonian and kicking operator for the system part, and the summation over $\alpha$ is omitted for simiplicity. In general order of inverse frequency, we can group the system-bath coupling part as
\begin{align}\label{seq:effectiveHamiltonian}
	D=D_V+H_B+\sum_{n=0}\sqrt{\gamma}\frac{Y_nB_n}{\Omega^{n}}+\mathcal{O}(\gamma),
\end{align}
\begin{align}
	iK(t)=iK_V(t)+\sum_{n=0}\sqrt{\gamma}\frac{Z_n(t)B_n}{\Omega^{n+1}}+\mathcal{O}(\gamma),
\end{align}
where $B_n\equiv \text{ad}_{H_B}^n[B]$ with $\text{ad}_{H_B}[B]\equiv\[H_B, B\]$ and $\vert Y_0\vert\sim \vert Z_n\vert\sim\mathcal{O}\(1\)$, $\vert Y_{n\geq1}\vert\sim\mathcal{O}\(\Omega^{-1}\)$. Notice that the system-bath coupling term includes multiple channels given by different order of comutator with bath Hamiltonian. Then the bath correlation function and spectral density among them are given by $J_{nm}(t-s)\equiv\langle B_n(t)B_m(s)\rangle$ and  $J_{nm}(\e)=\(\frac{\e}{\Omega}\)^{m+n}J(\e)$.
Applying Born-Markov approximation, one can obtain a time-independent Lindblad equation from Eq.(\ref{seq:effectiveHamiltonian})
\begin{align}
	\mathcal{L}_{D_V}\rho=-i[D_V+\Lambda_V,\rho]+\sum_n\(L_n\rho L_n^\dagger-\frac{1}{2}\left\{L_n^\dagger L_n, \rho\right\}\),
\end{align}
with jump operators given by
\begin{align}
	L_{n}=&\sqrt{\gamma}\sum_m\int_{-\infty}^\infty ds g_{nm}(s-t)e^{-iD_V(t-s)}Y_me^{iD_V(t-s)}\nonumber\\
	=&\sqrt{\gamma}\sum_m\sum_{ij}g_{nm}(\e_{ji})Y_{m,ij}\vert i\rangle\langle j\vert,
\end{align}
where $g_{nm}(\e)\sim \(\frac{\e}{\Omega}\)^{m+n}g(\e)$. 
Comparing this expression with the transition rate in FRE, detailed balance arises only when $g_{00}/J_{00}$ dominates, and the sub-leading correction occurs at order $\mathcal{O}\(\frac{g_{01}}{g_{00}}\)\sim\mathcal{O}\(\frac{\Lambda}{\Omega}\)$.
Thus with the leading order of jump operators, the steady state $\rho_S$ of the time-independent $\mathcal{L}_{D_V}$ tends to be the thermal state of $D_V$ approximately (ignoring the Lamb shift).  Then we check the effect of the periodic kicking from $iK(t)$ also under the Born-Markov approximation. The effective evolution of the system part can be obatined by tracing out the bath degree of freedom
\begin{align}\label{eq:kickingevoution}
	\mathcal{E}(t)\rho_S =& {\rm Tr}_B\(e^{-iK(t)}\rho_S\otimes\rho_Be^{iK(t)}\)\nonumber\\
	=&{\rm Tr}_B\(\rho_S\otimes\rho_B-\[iK(t), \rho_S\otimes\rho_B\]+\frac{1}{2}\[iK(t), \[iK(t), \rho_S\otimes\rho_B\]\]+\mathcal{O}(\Omega^{-3})\)\nonumber\\
	=&e^{-iK_V(t)}\rho_Se^{iK_V(t)}+\frac{\gamma}{\Omega^2}\sum_{n,m}\[\langle B_nB_m\rangle \(Z^nZ^m\rho_S-Z^m\rho_SZ^n\)+\langle B_mB_n\rangle\(\rho_SZ^mZ^n-Z^n\rho_SZ^m\)\]+\mathcal{O}(\Omega^{-3},\gamma^2),
\end{align}
where $\langle B_nB_m\rangle\sim\frac{\omega_c^{1+n+m}}{\Omega^{n+m}}$ for a spectral density with exponential cutoff. The first term in Eq.~\ref{eq:kickingevoution} represents the unitary kicking transformation on the system part and the second non-unitary term in Eq.~\ref{eq:kickingevoution} has the order $\mathcal{O}\(\frac{\gamma\omega_c}{\Omega^2}\)$, which can be dropped out under the high-frequency limit and weak-dissipation limit. Therefore, we conclude that a Floquet-Gibbs state can be realized when 
\begin{align}
	\Omega\gg\{\Lambda, \omega_c\},
\end{align}
with residual error $\mathcal{O}\(\frac{\Lambda}{\Omega}\)+\mathcal{O}\(\frac{\gamma\omega_c}{\Omega^2}\)$ for times up to the pre-thermal time $t_\ast= e^{-\mathcal{O}\(\Omega/{\rm max}\{\omega_c, \Lambda\}\)}$,
which is consistent with Eq.~\ref{eq:conditionforFGS}.

\end{document}